\documentclass[conference]{IEEEtran}
\IEEEoverridecommandlockouts
\usepackage{cite}
\usepackage{amsmath,amssymb,amsfonts}
\usepackage{graphicx}
\usepackage{textcomp}
\usepackage{xcolor}
\usepackage{subcaption}
\usepackage{todonotes}
\usepackage{bm}
\usepackage{algorithm}
\usepackage{algcompatible}
\usepackage{xcolor}
\usepackage{hyperref}

\newcommand{\blue}[1]{\textcolor{black}{#1}}

\def\BibTeX{{\rm B\kern-.05em{\sc i\kern-.025em b}\kern-.08em
    T\kern-.1667em\lower.7ex\hbox{E}\kern-.125emX}}
\begin{document}

\title{	Reinforcement Learning Approach for Mapping Applications to Dataflow-Based Coarse-Grained Reconfigurable Array}

\author{\IEEEauthorblockN{\textbf{Andre Xian Ming Chang\IEEEauthorrefmark{1}, Parth Khopkar\IEEEauthorrefmark{1}, Bashar Romanous, Abhishek Chaurasia, Patrick Estep,} \\
\textbf{Skyler Windh, Doug Vanesko, Sheik Dawood Beer Mohideen, Eugenio Culurciello}\\}
\IEEEauthorblockA{
\texttt{\{andrexianmin, pkhopkar, basharromano, achaurasia, pestep, swindh, dvanesko\}@micron.com} \\
\textit{Micron Technology, Inc.}\\
}
% \and
% \IEEEauthorblockN{Parth Khopkar*}
% \textit{Micron Technology, Inc.}\\
% \and
% \IEEEauthorblockN{Bashar Romanous}
% \textit{Micron Technology, Inc.}\\
% \and
% \IEEEauthorblockN{Abhishek Chaurasia}
% \and
% \IEEEauthorblockN{Patrick Estep}
% \textit{Micron Technology, Inc.}\\
% \and
% \IEEEauthorblockN{Skyler Windh}
% \textit{Micron Technology, Inc.}\\
% \and
% \IEEEauthorblockN{Doug Vanesko}
% \textit{Micron Technology, Inc.}\\
% \and
% \IEEEauthorblockN{Sheik Dawood Beer Mohideen}
% \and
% \IEEEauthorblockN{Eugenio Culurciello}
}

\maketitle
\begingroup\renewcommand\thefootnote{\IEEEauthorrefmark{1}}
\footnotetext{equal contribution}
\begingroup\renewcommand\thefootnote{}
\footnotetext{Code available at \url{https://github.com/micronDLA/RL_streaming_engine}}
\endgroup

\begin{abstract}
% The Streaming Engine (SE) is a Coarse-Grained Reconfigurable Array. % developed by Micron Technology.
% The SE provides programming flexibility and high-performance with energy efficiency.
% %A program is broken down into a set of one or more Synchronous Data-Flows (SDFs).
% A program is represented as a computation graph, where every instruction is a node.
% Each node needs to be mapped to the right slot and array in the SE to ensure the correct execution of the program.
% This creates an optimization problem with a vast and sparse search space.
% A manual mapping of the graph takes an infeasible amount of time by the programmer.
% Such manual mapping is impractical because it requires expertise and knowledge in the SE micro-architecture and reduces the search-space, thus, trading off optimization possibilities.
% In this work we propose a Reinforcement Learning(RL) based mapper to explore mappings and optimize them in an unsupervised manner using a reinforcement learning framework.
% This provides us an automated method that can produce mappings for programs quickly and searches for optimal mappings without the programmer's interference. 
% This tool also improves the usability of the SE device by encapsulating device configuration details.
The Streaming Engine (SE) is a Coarse-Grained Reconfigurable Array which provides programming flexibility and high-performance with energy efficiency.
An application program to be executed on the SE is represented as a combination of Synchronous Data Flow (SDF) graphs, where every instruction is represented as a node.
Each node needs to be mapped to the right slot and array in the SE to ensure the correct execution of the program.
This creates an optimization problem with a vast and sparse search space for which finding a mapping manually is impractical because it requires expertise and knowledge of the SE micro-architecture.
In this work we propose a Reinforcement Learning framework with Global Graph Attention (GGA) module and output masking of invalid placements to find and optimize instruction schedules.
We use Proximal Policy Optimization in order to train a model which places operations into the SE tiles based on a reward function that models the SE device and its constraints.
The GGA module consists of a graph neural network and an attention module. 
The graph neural network creates embeddings of the SDFs and the attention block is used to model sequential operation placement. 
We show results on how certain workloads are mapped to the SE and the factors affecting mapping quality.
We find that the addition of GGA, on average, finds 10\% better instruction schedules in terms of total clock cycles taken and masking improves reward obtained by 20\%.
% The implemented method is compared against evolutionary search and baseline.
\end{abstract}


\section{Introduction}
\label{sec:intro}
As Dennard scaling ends, big-data applications such as real-time image processing, graph analytics, and deep learning continue to push the boundaries of performance and energy efficiency requirements for computing system.
One solution to this challenge is to move compute closer to memory or storage for substantial energy savings of data movement.
We have developed an innovative Near-Data Computing (NDC) architecture that leverages the dramatic opportunities provided by the new CXL protocol \cite{sharma2020compute}.
NDC incorporates heterogenous compute elements in the memory/storage subsystem to accelerate various computing tasks near data.
One of these compute elements is the Streaming Engine (SE).

\blue{The SE is a Coarse-Grained Reconfigurable Array (CGRA) that is composed of interconnected compute tiles.
  The compute tiles are interconnected with both a Synchronous Fabric (SF) and an Asynchronous Fabric (AF).
  The SF enables neighboring tiles to be pipelined, forming an Synchronous Data Flow (SDF).
  The AF connects each tile with all the other tiles, as well as, the dispatch interface (DI), and memory interfaces (MI).}
Together, the SF and AF allow the tiles to efficiently execute high-level programming language constructs.
Simulation results of hand-crafted SE kernels have shown orders-of-magnitude better performance per watt on data-intensive applications than existing computing platforms.

\begin{figure}
  \centering
  \includegraphics[clip, width=\linewidth]{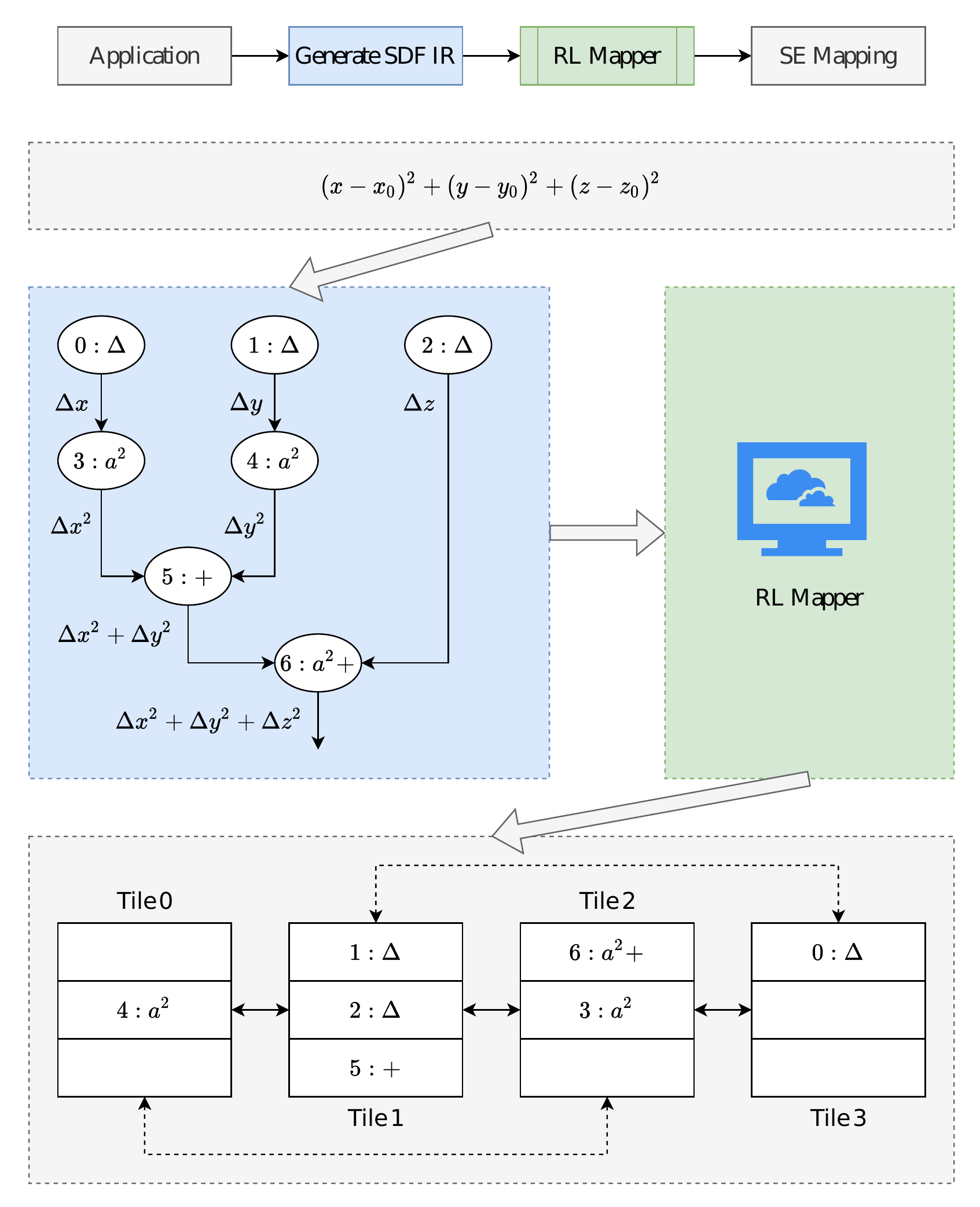}
  \caption{
    Example showing a mapping of a distance calculation function onto the SE device.
    The instruction execution latency is three clock cycles.
    The solid and dotted lines connecting the tiles, have one and two clock cycles transfer latency respectively.
    E.g., instruction \#0 is scheduled at (tile \#3, slot \#0) and its output is sent to instruction \#3.
    Instruction \#0 output is ready after three clock cycles plus two clock cycles to move data to tile \#1.
    Thus, instruction \#3 gets scheduled on (tile \#1, slot \#1).
  }
  \label{fig:se_example}
\end{figure}

A simple example that illustrates a program mapped to the SE is shown in \figurename~\ref{fig:se_example}.
The program in this example is a distance calculation function shown in Eq. \ref{eq:dist}.

\begin{equation}
  \label{eq:dist}
  D = \sqrt{(x - x_0)^2 +(y - y_0)^2 + (z - z_0)^2}
\end{equation}

To keep this example simple, we ignore the square root part of the equation.
The program is represented as an SDF graph as shown in \figurename~\ref{fig:se_example}.
Each operation in Eq. \ref{eq:dist} is an instruction that is mapped to a slot at a tile on the SE.

Since the output of the MS unit is connected to the input of the AL unit, instruction \#6 produces $\Delta z^2$ on the output on the MS unit and adds it to $(\Delta x^2 + \Delta y^2)$ at the AL unit.

Mapping the instructions from a program's SDFs onto the compute elements of the SE while adhering to architectural constraints is an NP-hard problem with a vast and sparse search space \cite{10.1007/3-540-69346-7_30}.
Constraints related to tile memory, synchronous dataflow, the use of delays to match timing requirements and more are necessary to ensure correct execution.
Creating the mappings manually or using brute force algorithms takes time and lots of effort even for the simplest of programs.
This process also adds assumptions that reduce the search space, trading off optimization possibilities.

In this work we propose a deep Reinforcement Learning (RL) method to explore and find optimal mappings in an unsupervised manner.
Using the Proximal Policy Optimization (PPO) method we train a neural network model to place instructions onto the SE tiles guided by a reward function in an environment that models the SE device and its constraints.
We also present Global Graph Attention (GGA) module that improves the baseline feedforward models by providing attention mechanism to the RL model.
Our RL method is combined with output masking, finetuning and sorted iteration order, which will be presented in the following sections.

The trained model is able to create valid mappings for the SE by learning about the structure of the problem domain.
On average addition of GGA finds 10\% better instruction schedules in terms of clock cycles for varying graphs complexities and different SE device configurations.

The key contributions of this paper are as follows:
\begin{enumerate}
  \item A Reinforcement Learning methodology that is able to map the instructions from a given application to the processing elements in the SE and has the potential to be reused across different applications.
  \item An analysis of different factors that impacts the quality of mappings obtained, such as attention module, output masking and iteration order.
\end{enumerate}

This work is motivated by the need to provide an improved SE toolset that lowers the SE usability barrier for a programmer.
In addition, the proposed mapper can be used along with other SE tools or assist in the manual mapping of applications to the SE by providing partial placement suggestions or tile configuration labels.
The RL mapper performs unsupervised learning and optimization allowing it to search a wide and sparse search space.
Each node in the SDF is an instruction that is mapped to a specific slot on a tile.

This line of research is inspired by recent work that used RL for chip placement \cite{mirhoseini2020chip}.
The problem requirement of mapping nodes of a SDF to available compute elements is similar to the problem of placing nodes of a chip netlist on a chip canvas.
% Due to the similarities between these problem domains, we theorize that our approach has the potential of being used for chip placement.
% This hypothesis will be investigated in future work.

The rest of this paper is organized as follows:
Section \ref{sec:relatedwork} discusses related work.
Section \ref{sec:background} provides the necessary background on the SE device architecture needed to understand the details of the RL approach.
Section \ref{sec:rlmapper} presents the proposed RL approach for mapping applications to the SE.
Section \ref{sec:results} discusses results, and Section \ref{sec:conclusion} concludes the paper along with a discussion on future work.

\section{Related work}
\label{sec:relatedwork}
\subsection{CGRA}.
CGRAs are a heavily researched architectural paradigm with a long history and can provide an excellent balance of high-performance compute, memory bandwidth, and area and energy efficiency \cite{theodoridis2007survey}.
Due to such advantages, CGRAs are currently enjoying a resurgence in interest, not just in the research realm \cite{prabhakar2018plasticine}, but also commercially \cite{morgan2018intel, nicol2017coarse, vissers2019versal}.
Research compilers targeting CGRAs are available (for example, \cite{adriaansen2016code, chin_cgra-me_2017, mei2003exploiting, prabhakar2018plasticine}) but they are limited in quality and code coverage. 
Industry-strength compilers such as Clang/LLVM do not provide official support for CGRA-like architectures.
For further information on the various taxonomy of CGRAs architectures and design, we refer the reader to \cite{liu_survey_2019, tehre_survey_2012}.

Mapping various programming constructs onto CGRAs is an extensive research topic.
The  mapping of large and irregular loops onto CGRAs is analyzed in \cite{zhao_towards_2020} which proposed paying more attention to temporal mapping than spatial mapping. 
The proposed temporal mapping utilizes a buffer allocating heuristic with constraint on computations and interconnection resources. 
The proposed spatial mapping uses a backtracking and reordering mechanism with greedy algorithm with the goal of minimizing the Initiation Interval (II).
A method for mapping SDFs to a CGRA is introduced in \cite{li_chordmap_2022}.
The proposed method relies on modulo-scheduling to \blue{provide} a spatio-temporal mapping.
ChordMap creates a schedule and partitions the SDF and CGRA, then performs spatio-temporal mapping of every kernel instance in each partition.
ChordMap operates on three levels of parallelism: application-level, kernel instances-level, and instruction-level parallelism.
%Parallel execution and high throughput are enabled by by time slicing the CGRA resources.
In CGRA-ME\cite{chin_cgra-me_2017}, a CGRA device model is built using Module Routing Resource Graph (MRRG) \cite{mrrg}. 
The parser converts an optimized C-language benchmark into a Data Flow Graph (DFG) using LLVM compiler framework. 
Each operation in the DFG is mapped to a functional unit in the MMRG using simulated annealing. 
The routing between inputs and outputs of each operation is selected using PathFinder-like algorithm \cite{pathfinder}.

SE differentiates from others as the first CGRA in a near-data computing architecture.
SE also provides asynchronous messaging as a first-class programming construct along with an SDF. 
%A key challenge with SE applications is efficient compilation of high-level language code (for example, in C/C++) to executable program. 

\subsection{Reinforcement Learning}
RL is widely used to tackle unsupervised optimization problems.
\blue{Typically in RL, a policy is a function that outputs the action that an agent should take in a system given the current state of the system.}
It has been applied in chip placement \cite{mirhoseini2020chip}, workload distribution \cite{Mirhoseini_placementRNN, addanki2019placeto, zhou2019gdp}, compiler optimizations \cite{Zhou_compileGNN} and other decision based tasks \cite{kormushev2013reinforcement, ZophL16_NASRL}. 
Alternative optimization algorithms include evolutionary strategies \cite{Zhichao_ESNAS} and bayesian optimization \cite{shi2020learned}. 
An advantage of an RL approach is that it is able to learn from a collection of programs and reuse previous data for new programs by training a Deep Learning model \cite{zhou2019gdp}.
\blue{When RL is used in a deep learning paradigm, the policy is represented by neural networks.}

Recurrent Neural Networks (RNN) \cite{hochreiter1996lstm} were previously a popular approach to process a sequence of nodes from a graph representation. 
Recently, Graph Neural Networks (GNN) \cite{gori2005new} have shown success in processing structured data without needing the preprocessing required for RNNs.
GNNs are widely used for tasks involving graph processing \cite{Zhou_compileGNN, zhou2019gdp}. 
Attention modules have sometimes been used in literature to supplement the embeddings created by GNNs to further improve results \cite{addanki2019placeto}.

In \cite{mirhoseini2020chip}, an RL based method for chip placement is presented where a graph neural network is used to create embeddings from a netlist graph and then passed through an actor model to get placements on a chip canvas.
The training is done using the PPO approach and one component from the netlist is placed at each step until all required components have been placed.
The actor network in this method is composed of deconvolution layers which are more computationally intensive than our approach. 
An RL based method, proposed in \cite{wu_core_2020}, that uses a Deep Deterministic Policy Gradient (DDPG) method for component placement in multi-chip many-core systems.
A review of various machine learning approaches \blue{is} presented in \cite{khailany_accelerating_2020} which including reinforcement learning for chip design.
The work proposed in \cite{zhou2019gdp} presents a combination of graph neural network and transformer-XL model to place operations in dataflow graphs on suitable devices.
In each of these works, the aim is similar to ours i.e., to reduce the manual labor and domain expertise required to produce mappings.

% Our input is a combination of computation graph, SE device state and the node to be placed at a particular time step. 
% Instead of only feeding our actor model with embedding from graph neural network, we combine the graph neural network embedding with information from the SE device and a representation of node that is going to be placed. 
% A configuration is also added to guide the model to optimize different goals.

A key difference between our work and \cite{zhou2019gdp} is that our strategy is to place one node at a time, instead of generating a complete assignment per iteration. 
This allows our model to break down the placement problem into sub-problems and also allows the framework to start from a different initial configuration. 
For example, if some nodes are already placed by some other algorithm such as brute force, the RL mapper can place the remaining nodes. 
This approach also allows us to obtain more data during the sampling phase which consists of partial assignments, instead of only saving one training sample for an entire sequence of nodes.

\section{Background on the SE}
\label{sec:background}
\blue{The SE is a coarse-grained fabric, shown in \figurename~\ref{fig:sub-se}, composed of compute elements or tiles.
These tiles are interconnected with an SF, allowing data to traverse from one tile to another without queuing.
This SF allows many tiles to be pipelined together to produce a continuous data flow through SIMD arithmetic operations.
Each tile, interconnects local memory, multiplexers, a Multiply/Shift (MS), and an Arithmetic/Logic (AL) SIMD capable units as shown in \figurename~\ref{fig:sub-tile}.} 
The tiles are also interconnected with an AF that allows synchronous domains of compute to be bridged by asynchronous operations.
These asynchronous operations include initiating SDF operations, transferring data from one SDF to another, accessing system memory (read and write), and performing branching and looping constructs.

\begin{figure} [ht]
  \begin{subfigure}{.5\textwidth}
    \centering
    \includegraphics[trim=118 310 180 110, clip, width=0.7\linewidth]{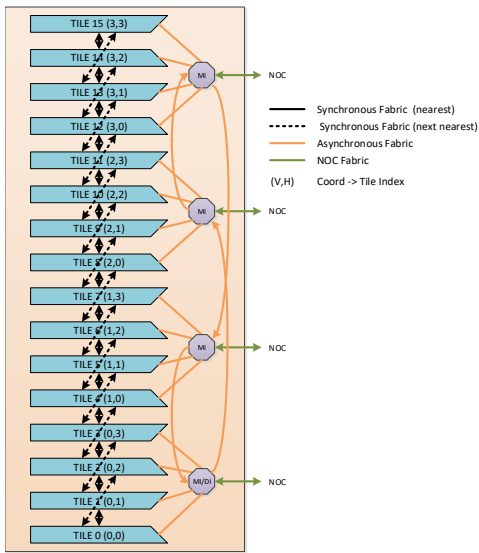}
    \caption{SE device}
    \label{fig:sub-se}
    \end{subfigure}
  \begin{subfigure}{.5\textwidth}
  \centering
  \includegraphics[trim=300 100 8 50, clip, width=0.5\linewidth]{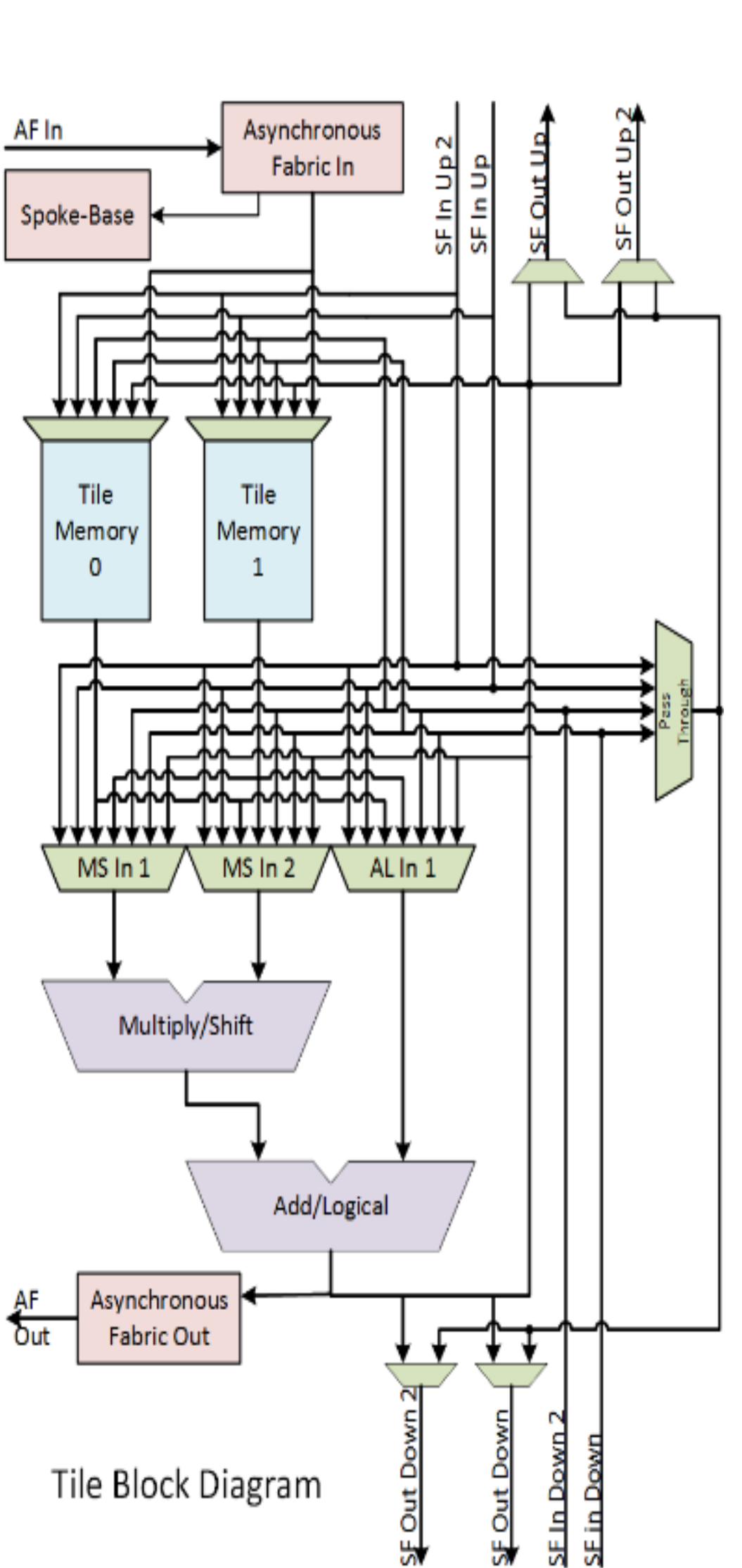}
  \caption{SE tile}
  \label{fig:sub-tile}
  \end{subfigure}
  \caption{
    \figurename~\ref{fig:sub-se} shows layout of the SE device and dataflow paths between tiles. \figurename~\ref{fig:sub-tile} is a detailed diagram of a tile in the SE device. 
    Note the two tile memory units, the MS and AL units, and the four SF inputs and four SF outputs.
  }
  \label{fig:se_diagram}
\end{figure}

As shown in the \figurename~\ref{fig:se_diagram}, tiles are interconnected in a $1 \times 16$ configuration.
The SF is used to connect a tile to another tile one hop above it, a tile two hops above it, a tile one hop below it and a tile two hops below it.
Information is transferred over the SF with a deterministic latency. 
Each tile acts independently, streaming data through internal memory and MS/AL units to other tiles over the SF and AF.
The tiles use the AF to communicate between synchronous domains, send loads and stores to memory through the MI, and receive commands from the host through the DI to initiate work on the SE.
Information is transferred over the AF with a non-deterministic latency.

\subsection{Synchronous Fabric Interface}
All tiles that participate in a synchronous domain act as a single pipelined data path.
\blue{The SDF's entry tile is defined as the tile that executes the first instruction of the SDF.
It is responsible for initiating a thread of work through the pipelined tiles at a predefined cadence, referred to as the Spoke Count or II.
E.g., For $II = 3$, as in the example in \figurename~\ref{fig:se_example}, then the entry tile can initiate work every third clock cycle.}
%The synchronous fabric provides both data and control information.

\subsection{Asynchronous Fabric interface}
The Asynchronous Fabric is used to perform operations that occur asynchronously to a synchronous domain.
Each tile contains an interface to the AF.
AF messages can be classified as either data messages or control.
Data messages contain a SIMD width data value that is written to one of the two tile memories.
Control messages are for controlling thread creation, freeing resources, or issuing external memory requests.

\subsection{Tile Base}
\blue{The tile base contains data structures that are used to initiate an SDF.}
%It receives messages from the SF interface to configure state and set up the necessary conditions to start a synchronous flow.
%It initiates a flow into the SF when the required conditions are met.
%It also contains control logic to manage iterations of loops.
\blue{If the tile was an entry tile of an SDF, then the tile base launches a new thread every II.}
The II allows the base logic to launch new threads or continue previously launched threads when all resources needed for execution are available.

\subsection{Tile Memory}
Each tile contains two memories.
Each are the width of the data path (512-bits), and the depth will be in the range of 512 to 1024 elements.
The tile memories are used to store data required to support data path operations.
The stored data can be constants loaded as part of the program's arguments, or variables calculated as part of the data flow.
The tile memory can be written from the AF as either a data transfer from another SDF, or the result of a load operation initiated by another SDF.
%The tile memory is only read via synchronous data path instruction execution.

\subsection{Instructions RAM}
Each tile has an instruction RAM has multiple entries to allow a tile to be time sliced, performing multiple, different operations in a pipelined synchronous domain.
%Additionally, each time slice could conditionally perform different instructions depending on previous tile time slice data dependent conditional operations. 

\subsection{Spoke RAM}
The Spoke RAM has multiple entries to reconfigure the tile at each time slice (clock cycle).
The number of active spoke RAM slots in a tile is equal to the II of that tile.
A couple of relevant configurations are: 

\begin{itemize}
  \item Which of the four SF inputs, feedback from the output of the tile's MS/AL unit, or the tile base is the master input.
  \item Which of the four SF outputs are used to send the output of the MS/AL unit to another tile or tiles using SF.
\end{itemize}

The Spoke RAM iterates over its slots using a counter that modulo counts from zero to II minus one and back to zero.
%Using different spoke counts on different tiles can be a powerful mechanism that allows the number of slices required by an inner loop to determine the performance of an application.
The proposed RL mapper provides the II and the configuration of each spoke RAM slot, including which tile instruction will be active, on each utilized tile.

\subsection{Programming the SE}
The SE programmer breaks down the desired application into a set of one or more SDFs.
An SDF is marked by instructions sending and receiving data in deterministic latency.
All operations with nondeterministic latencies, e.g., load/store requests to the MI, mark either the beginning and/or ending of an SDF.
We have written an in-house parser that enables the programmer to express the program in terms of SDFs using our defined assembly language.
The parser produces an Intermediate Representation (IR) that is used as an input to our proposed mapper.

The IR is a graph representation of the program which consists of SDF subgraphs. 
Each SDF is a disconnected component of the IR graph. 
Here every instruction \blue{is} a node.
Each node represents an instruction that needs to be placed onto a SE tile at the proper time-slice such that when all nodes are placed, the program executes correctly. 
In \figurename~\ref{fig:se_example}, the edges of the graph represent the data dependencies of the instructions. 
The nodes also contain information about the variables that need to be present in tile memories during instruction execution. 
The instruction intended to execute at a specific time-slice (clock cycle) will only execute when the master input configured in the corresponding Spoke RAM entry receives a valid control message.
%The tiles can pass data to their neighbors and each tile can be configured with a different number of initiation intervals (II). 
%Each II can be allocated to run a single instruction during program execution. 
%After one cycle, the tiles will move on to the next II, with execution returning to first II after the last. 
%All tiles run the instruction in the same II in parallel. 
%Thus, whenever possible, independent instructions should be mapped to different tiles and same II to exploit parallelism. 
%The first operation of each disconnected component of the computation graph (one component representing a synchronous flow) needs to be placed on different tiles. 

\subsection{SE Mapping Constraints}
The SE hardware imposes constraints that the RL mapper must adhere to in order to for it to produce a valid mapping. These constraints are:
\begin{enumerate}
  \item Instructions that share one or more tile memory variables must be placed on the same tile.
  \item No two or more instructions that start an SDF can share a tile.
  \item No two or more instructions that are siblings in the SDF can share a tile.
\end{enumerate}
The SE device configuration and its constraints are modeled in a simulation environment and a reward function is used to determine the quality of placements obtained.
To enforce these constraints, we explore two methods in Section \ref{subsec:output_masking}.
The first method is to give a negative reward when a constraint is violated and the second is creation of a mask on the invalid actions so that the network only outputs valid actions.
In both cases, placing instructions sub-optimally leads to a reduced reward whereas placements that optimally reduce total execution time of the graph are rewarded. 

\section{Method}
\label{sec:rlmapper}
\subsection{Reinforcement learning}

We present an RL based framework to explore and optimize the mapping of instructions for a given application to the SE, guided by a reward function that informs the mapping algorithm about the quality of the produced mappings at each step. In this section, we give an overview of the methods used along with the formulation required for detailed description of the various components of the RL approach.

\subsection{Overview of RL Methodology}
PPO \cite{schulman2017proximal} is an RL method that is  widely used for continuous and discrete action problems. 
It trains an actor and a critic model (represented as neural networks).
The actor model is trained to produce actions (node placements in our case) from sample states obtained during simulation and the critic model is trained to match the sampled rewards from the actions produced by the actor using a surrogate loss function. 
This sample and train process is repeated over various iterations. 
In this manner, the problem search space is explored using the reward function as a heuristic.
Refer to algorithm Alg.~\ref{alg1} for the complete RL training process.
Our SE mapping task is formulated as a discrete action problem where the goal is to place one node at each time step. 
Samples of SE state, actions performed, and rewards obtained at each time step are collected after executing the actions provided by the actor model in simulation. 
These samples are stored in a buffer and are used as data to train the models.
\figurename~\ref{fig:ppo} shows the overall RL framework. 

The key components of our RL method are as follows:
\begin{itemize}
  \item States: The state is $s$ represented by a concatenation of current SE state, an embedding of the whole computation graph and the selected node to be placed next.
  \item Action: The action $a$ consists of the node to be placed along with the tile and spoke location it is to be placed at.
  \item Reward function: The reward obtained is based on the number of clock cycles taken for a node to finish executing after its predecessors finished executing.
  % \item Transition function: The transition function gives the probability distribution over next states given the current state and the action to be performed.
\end{itemize}

\begin{figure}[tb]
  \centering
  \includegraphics[trim=15 30 20 20, clip, width=\linewidth]{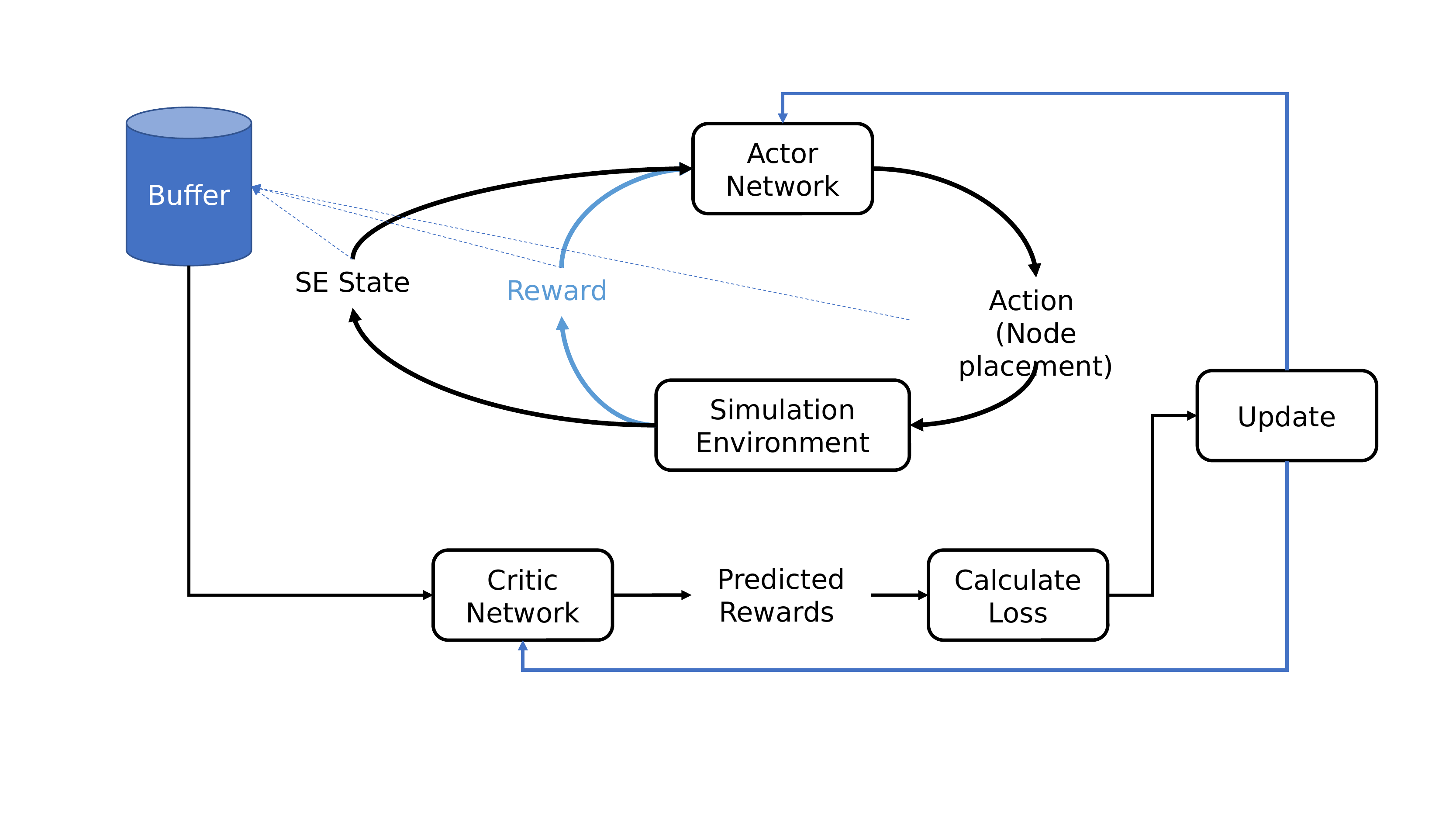}
  \caption{Diagram of the RL framework showing the role of actor and critic networks during RL training. }
  \label{fig:ppo}
\end{figure}

\subsection{State Representation}
The state $s$ is a vector of size $|TS| + f + 1$ where $TS$ is the set of tile slices in the Streaming Engine with $TS=T \times S$ and $f$ is the length of graph features. 
Here $T$ and $S$ are sets of all tiles and slots in the SE respectively. 
The number of slots available in each tile is equal to number of Initiation Intervals.

\subsection{Action Representation}
The action $a$ at each step is a tuple \((n,t,s)\) where $n \in N$ (the set of all nodes) is the node to be placed and $t \in T$ and $s \in S$ are respectively the tile and slot at which the node $n$ is to be placed. 
We place nodes in a topological order which ensures that all of a node's predecessors have been placed before the node. 
We do not make selecting the node to be placed a part of the learning problem since it increases the complexity of the learning process and makes training harder.

\paragraph*{\blue{Masked Actions}}
In order to enure that the network only outputs valid actions, we determine a binary mask over all possible actions and set the value of logits (unnormalized outputs of last layer of the neural network) corresponding to invalid actions to $-\infty$ in the actor network. 
This in turn sets the probability of sampling invalid actions to $0$, ensuring that we never take an invalid action.
Finally, we only calculate entropy on valid actions, making sure that our algorithm maximizes exploration only for valid actions in a given state.

\subsection{Reward Function}
Our goal in this work is to get mappings that are optimal in terms of total clock cycles taken, and for this purpose, the reward that is obtained at each time step is the difference between the clock cycle at which the current node to be placed is ready (its ready time) and the ready time of its predecessor. 
$R_n$ is the reward obtained for placing node $n$ and $t_n$ is its ready time. 
The function $p(n)$ gives the predecessor of $n$ in an SDF graph. 
If the current node can't be placed because of the constraints (all values in the mask vector $m$ are zero), then a high negative reward $-\lambda$ is given. If a node has more than one predecessor, then the node with the later ready time is chosen for the purpose of determining reward value.
\[
  R_n =
  \begin{cases}
    -\lambda,& m_i = 0, \, \forall \, i \in T \times S \\
    t_n - t_{p(n)}, & \text{otherwise}
    
  \end{cases}
\]

\begin{figure*}[tb]
  \centering
  \includegraphics[width=\textwidth]{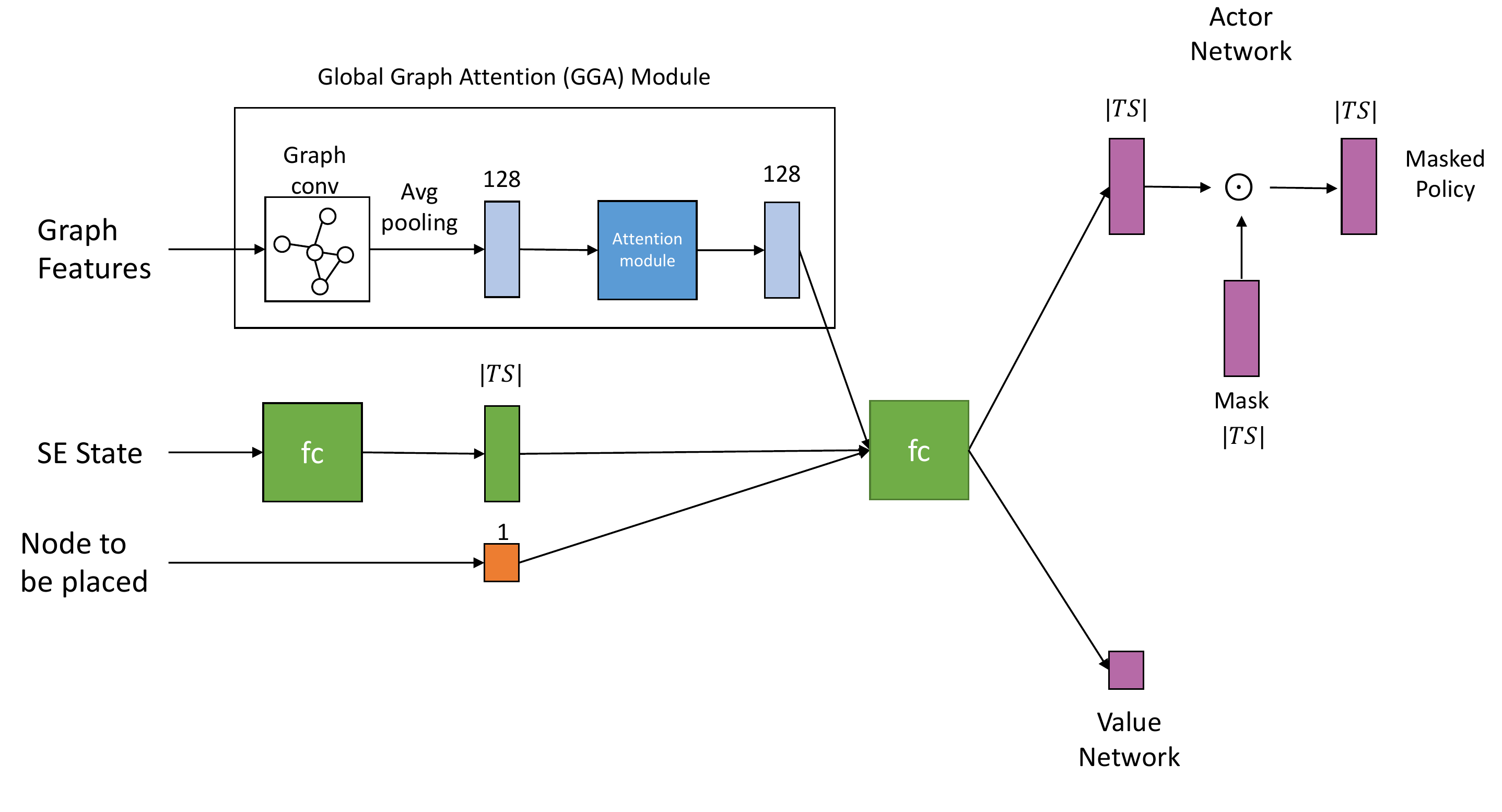}
  \caption{Actor and critic model architecture. 
  GNN is used to process the SDF graphs (static data). 
  Attention module is used to determine relative importance of nodes that are relevant to the one currently being placed. 
  The embedding created from dynamic data is combined with static data embedding. 
  A final MLP model is used to generate actions. 
  Actions are masked to ensure only valid actions are produced. }
  \label{fig:model}
\end{figure*}

\subsection{Model design}
The architecture for actor and critic models is shown in \figurename~\ref{fig:model}.
The input is separated in two categories: static and dynamic data. 
Static data is information that doesn't change as nodes are being placed and includes the graph features from the SDFs that encapsulate node dependence based on SE constraints.
Device state (including placed nodes) and node to be placed are dynamic data that change during placement.

\subsection{Global Graph Attention Module}

A GNN is used to process the IR graph where each node contains a feature vector that is initialized with a combination of tile memory constraints and node data dependencies. 
The input IR for an application is the same during placement of each node in the SDFs for that application.
After placing all nodes in an episode, a different IR graph for another application can be fed into the RL model to train it on a collection of application tasks. 

The Global Graph Attention Module is the combination of the GNN and attention modules.
The GNN is composed of two layers of graph convolution \cite{wu2019simplifying} and a graph average pool layer which produce a vector embedding.
An attention module is applied to the embedding produced by the GNN to highlight relevant components for the current node to be placed. 
We evaluated two attention module implementations. A transformer encoder layer \cite{vaswani2017attention} and a modified Position Attention Module (PAM) from \cite{fu2019dual}, in which the positions represent different node embeddings.
In our experiments, both produced similar results. 
Having a higher number of transformer encoder layers improves the best reward obtained but it is achieved at a cost of more parameters, making training more expensive.

The selected node and SE state change for each iteration over all nodes in the SDF graph. 
This dynamic data is fed into a Fully Connected (FC) layer model to create another embedding to represent the current state. 
The embeddings from dynamic data and static data are concatenated and fed to a MultiLayer Perceptron (MLP) model with three FC layers to obtain placements. 
Invalid actions are masked before being sent to the SE simulation environment. 
Output masking to filter invalid placements has previously been shown to be effective for RL in game environments \cite{Shengyi_mask}. 
We see the addition of GGA module provides improved reward and sample efficiency for various applications and different device configurations as shown in Section \ref{sec:GGA_result}.

% \begin{algorithm*}
%   \caption{\textcolor{red}{RL Mapping Algorithm}}
%   \label{alg:mapping}
%   \begin{algorithmic}[1]
%     \State Input: $\theta_0$: Initial actor network parameters, $\phi_0$: Initial value network parameters, Graph $\mathcal{G}$ which is to be placed
%     \For{k=0,1,2,...}
%       \State Collect set of placements $\mathcal{D}_k$ by running policy $\pi(\theta_k)$.
%       \State Compute advantage estimates $\mathit{A}$ using value function $\mathit{V}$
%       \State Update policy $\pi(\theta_k)$:
%       $\theta_{k+1} = \arg \max_{\theta} \underset{s,a \sim \pi_{\theta_k}}{{\mathrm E}}\left[
%         \min\left(
%           \frac{\pi_{\theta}(a|s)}{\pi_{\theta_k}(a|s)}  A^{\pi_{\theta_k}}(s,a), \;\;
%           \text{clip}\left(\frac{\pi_{\theta}(a|s)}{\pi_{\theta_k}(a|s)}, 1 - \epsilon, 1+\epsilon \right) A^{\pi_{\theta_k}}(s,a)
%           \right)\right],$

%     \EndFor
%   \end{algorithmic}
% \end{algorithm*}

\begin{algorithm*}
  \caption{RL Mapper Training Algorithm}
  \label{alg1}
\begin{algorithmic}[1]
  \STATE Input: $\theta_0$: Initial actor network parameters, $\phi_0$: Initial value network parameters, Graph $\mathcal{G}$ which is to be placed
  \FOR{$k = 0,1,2,...$}
  \FOR{$i = 0,1,2,...$}
  \FOR{$\text{node} \, n \in \mathcal{G} \, \text{in topological order}$}
    \STATE Get placement mask $m$ for $n$
    \STATE Get placement for $n$ by running policy $\pi(\theta_k, m)$
    \STATE Store node placements in buffer ${\mathcal D}_k$.
  \ENDFOR
  \ENDFOR
  \STATE Compute discounted rewards $\hat{R}_t$.
  \STATE Compute advantage estimates, $\hat{A}_t = \hat{R}_t - \hat{V}_t$ using current value function $V_{\phi_k}$.
  \STATE Update the policy by maximizing the PPO-Clip objective:
      \begin{equation*}
        \theta_{k+1} = \arg \max_{\theta} \frac{1}{|{\mathcal D}_k| T} \sum_{\tau \in {\mathcal D}_k} \sum_{t=0}^T \min\left(
            \frac{\pi_{\theta}(a_t|s_t)}{\pi_{\theta_k}(a_t|s_t)}  A^{\pi_{\theta_k}}(s_t,a_t), \;\;
            g(\epsilon, A^{\pi_{\theta_k}}(s_t,a_t))
        \right),
      \end{equation*}
      via stochastic gradient ascent with Adam where
      \begin{equation*}
        g(\epsilon, A) = \left\{ 
          \begin{array}{ll}
          (1 + \epsilon) A & A \geq 0 \\
          (1 - \epsilon) A & A < 0.
          \end{array}
          \right.
      \end{equation*}
  \STATE Fit value function by regression on mean-squared error:
      \begin{equation*}
      \phi_{k+1} = \arg \min_{\phi} \frac{1}{|{\mathcal D}_k| T} \sum_{\tau \in {\mathcal D}_k} \sum_{t=0}^T\left( V_{\phi} (s_t) - \hat{R}_t \right)^2,
      \end{equation*}
      via gradient descent algorithm.
  \ENDFOR
\end{algorithmic}
\end{algorithm*}

\section{Results}
\label{sec:results}

The aim of this section is to evaluate the RL framework for mapping applications to the SE.
We analyze the effects that different components of the RL model such as GGA module and training strategies such as node iteration ordering and output masking have on the quality of placements generated. We also analyze the results of pre-training the RL model on multiple graphs and then fine-tuning it on a desired application.

The implemented RL approach was able to successfully map different applications such as vector add, distance calculation function and Fast Fourier Transform (FFT).
FFT is widely used in several areas such as digital signal and image processing, pattern recognition, solving partial differential equations, error-correcting codes, and many others \cite{814659}.
FFT has time complexity of $\mathcal{O}(N\log{N})$ and two nested loops.
The outer loop and inner loop have time complexity of $\mathcal{O}(\log{N})$ and $\mathcal{O}(N)$ respectively.
The inner loop is targeted for acceleration.
The inner loop code is optimized and broken down into four SDFs which are shown in \figurename~\ref{fig:ifft_graph}.
The SE device configuration for the following experiments used 16 tiles and a maximum slot count of six.
Other SE device configurations are possible.

\begin{figure}[tb]
  \centering
  \includegraphics[width=\linewidth]{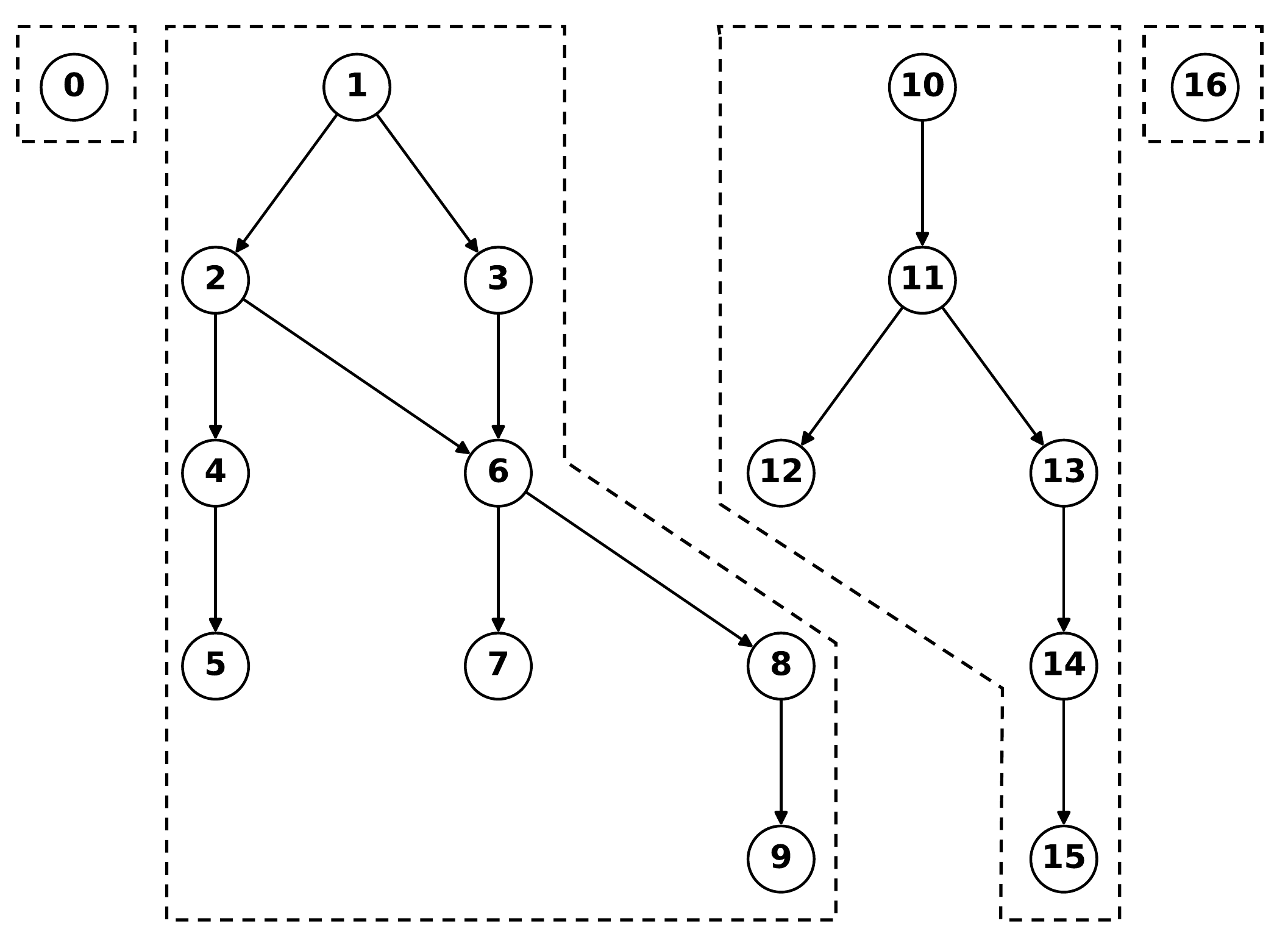}
  \caption{IR for FFT application consists of 4 SDFs. The arrows show data dependence between nodes. Tile memory dependencies have not been shown for the sake of simplicity.}
  \label{fig:ifft_graph}
\end{figure}

\begin{figure}[tb]
  \centering
  \includegraphics[width=\linewidth]{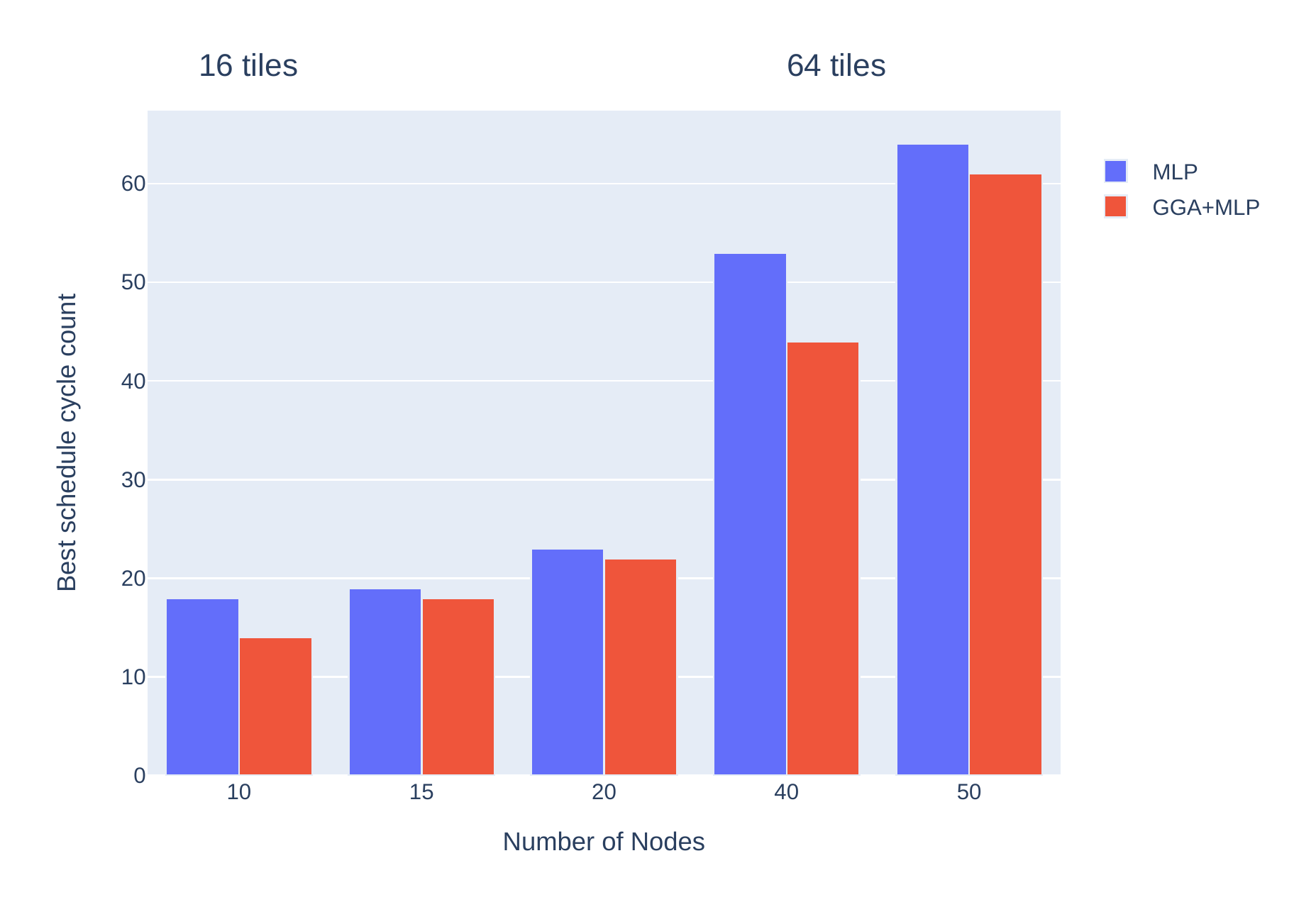}
  \caption{Cycle count for running all nodes in the best mapping given by RL model over 50,000 epochs.
    PPO baseline MLP model and GGA model were evaluated with computation graphs with increasing number of nodes.
    A larger device configuration with 64 tiles was used for experiments with IR graphs with 40 and 50 nodes. }
  \label{fig:nodes_graph}
\end{figure}

\begin{figure}[tb]
  \centering
  \includegraphics[width=\linewidth]{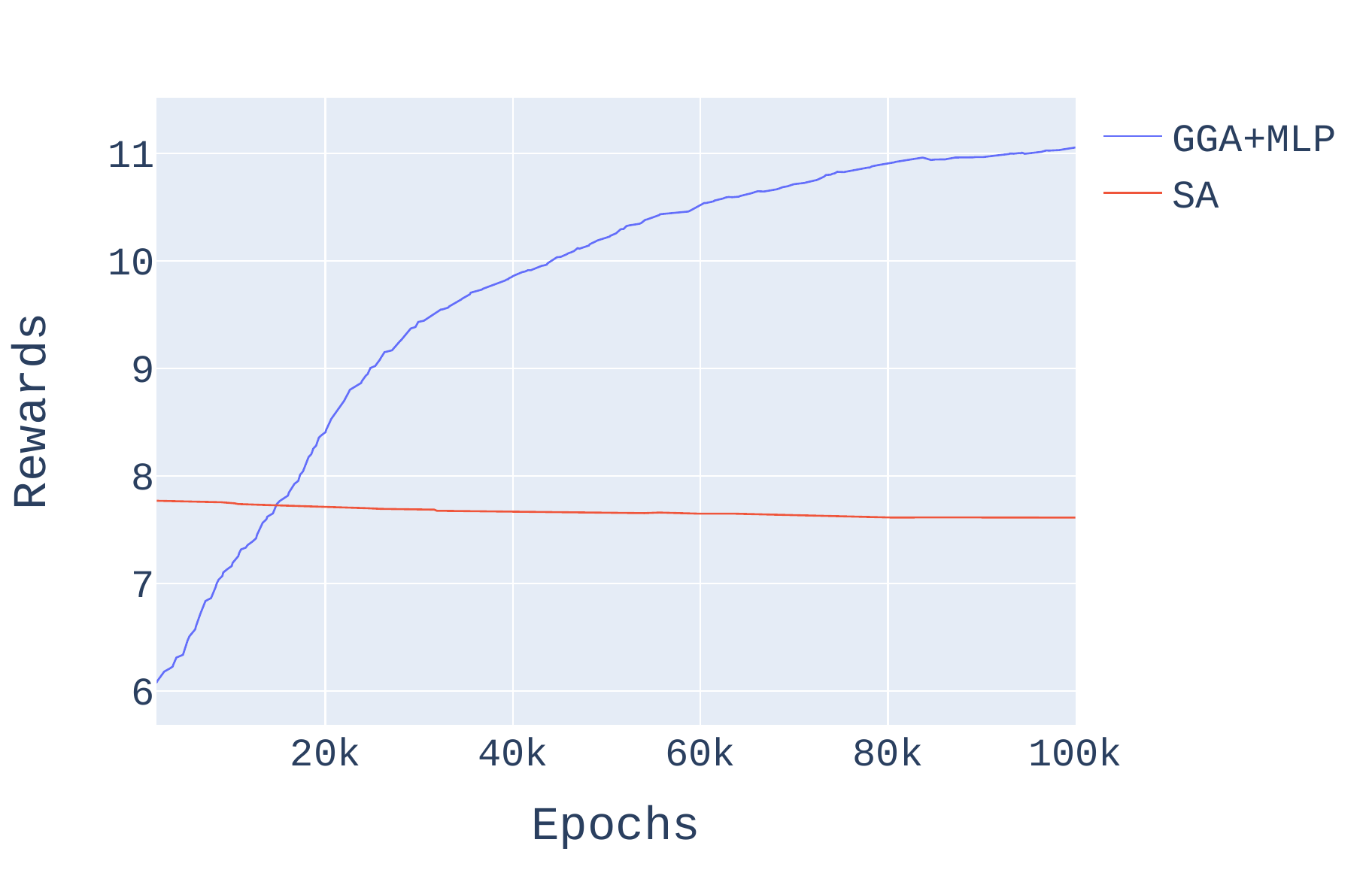}
  \caption{\blue{Comparison between training RL model (blue) versus Simulated Annealing (SA) method (red). }}
  \label{fig:ifft_sa}
\end{figure}

\begin{figure}[tb]
  \centering
  \includegraphics[width=\linewidth]{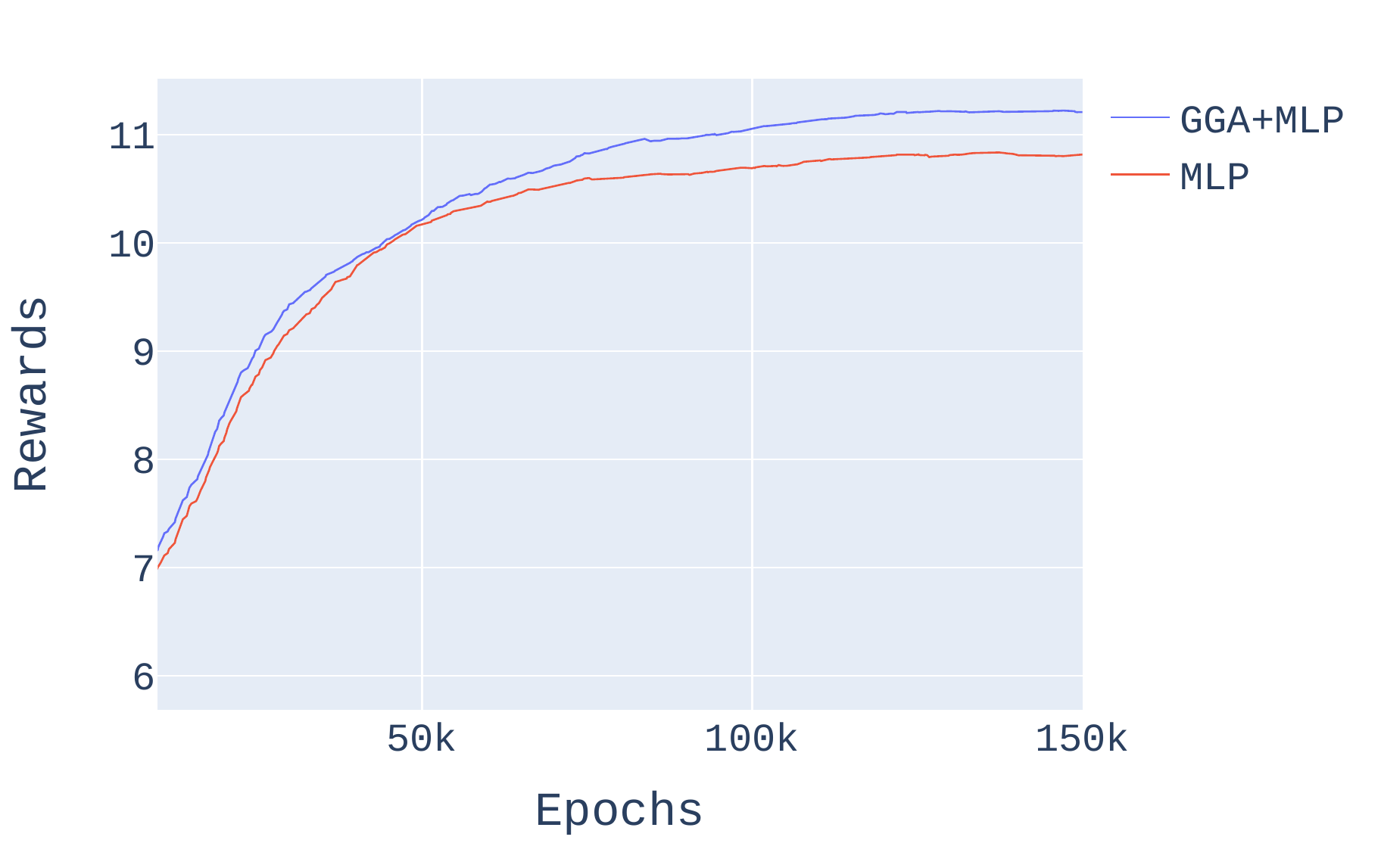}
  \caption{Effect of using Global Graph Attention (GGA) module for mapping the FFT application on device with 16 tiles.
    GGA module provides better sample efficiency and higher reward after training. }
  \label{fig:ifft_rewards}
\end{figure}

\begin{figure}[tb]
  \centering
  \includegraphics[width=\linewidth]{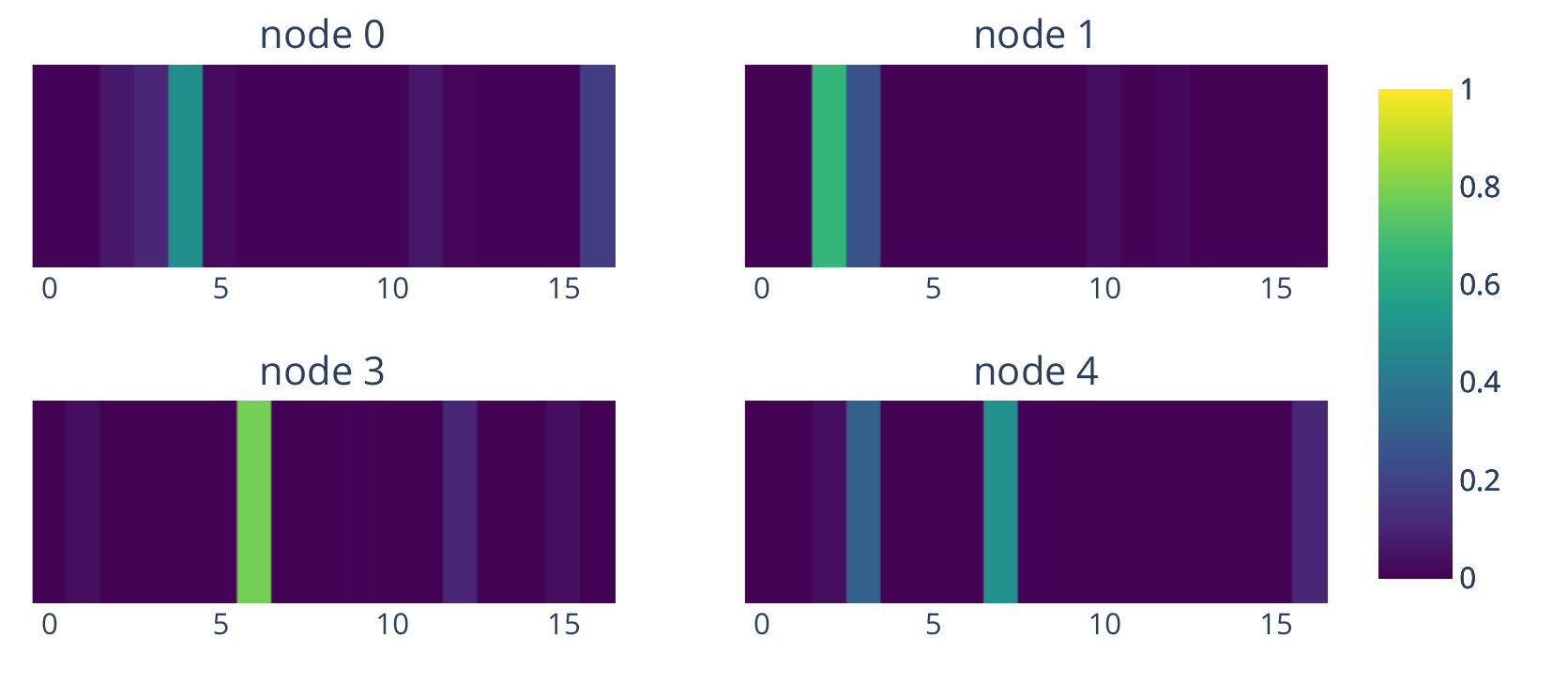}
  \caption{Attention scores from transformer module when placing certain nodes for the FFT application. }
  \label{fig:ifft_attention}
\end{figure}

\begin{figure}[tb]
  \centering
  \includegraphics[width=\linewidth]{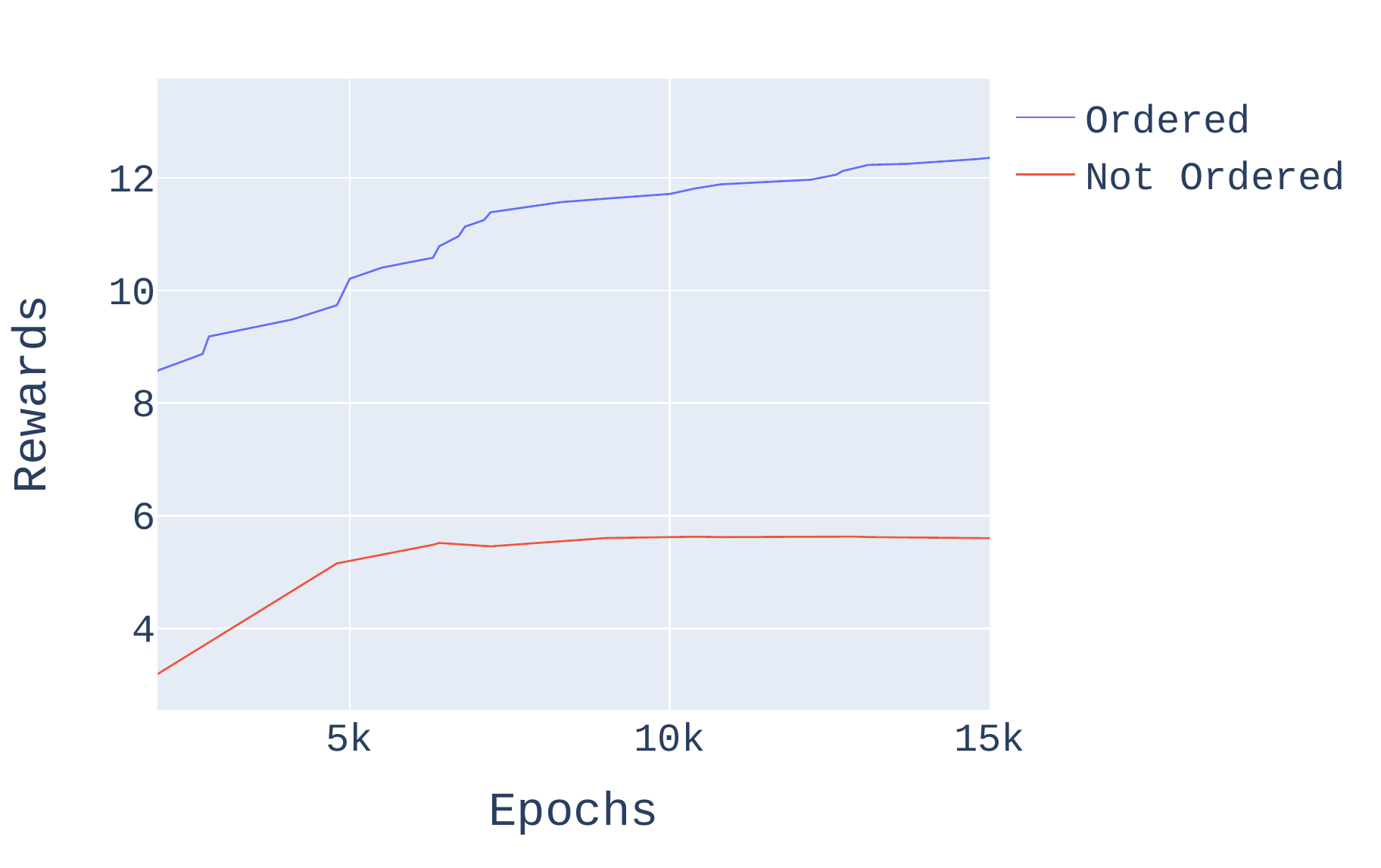}
  \caption{Iterating nodes in topological ordering results in higher reward and eases the placement task.
    The blue line is the reward curve for when nodes were randomly selected for placement.
    The red line is the reward curve for when nodes were iterated upon in topological order.
    The task involved is of placing 15 nodes of an IR graph onto a device with 16 tiles.}
  \label{fig:ordered_placement}
\end{figure}

\begin{figure}[tb]
  \centering
  \includegraphics[width=\linewidth]{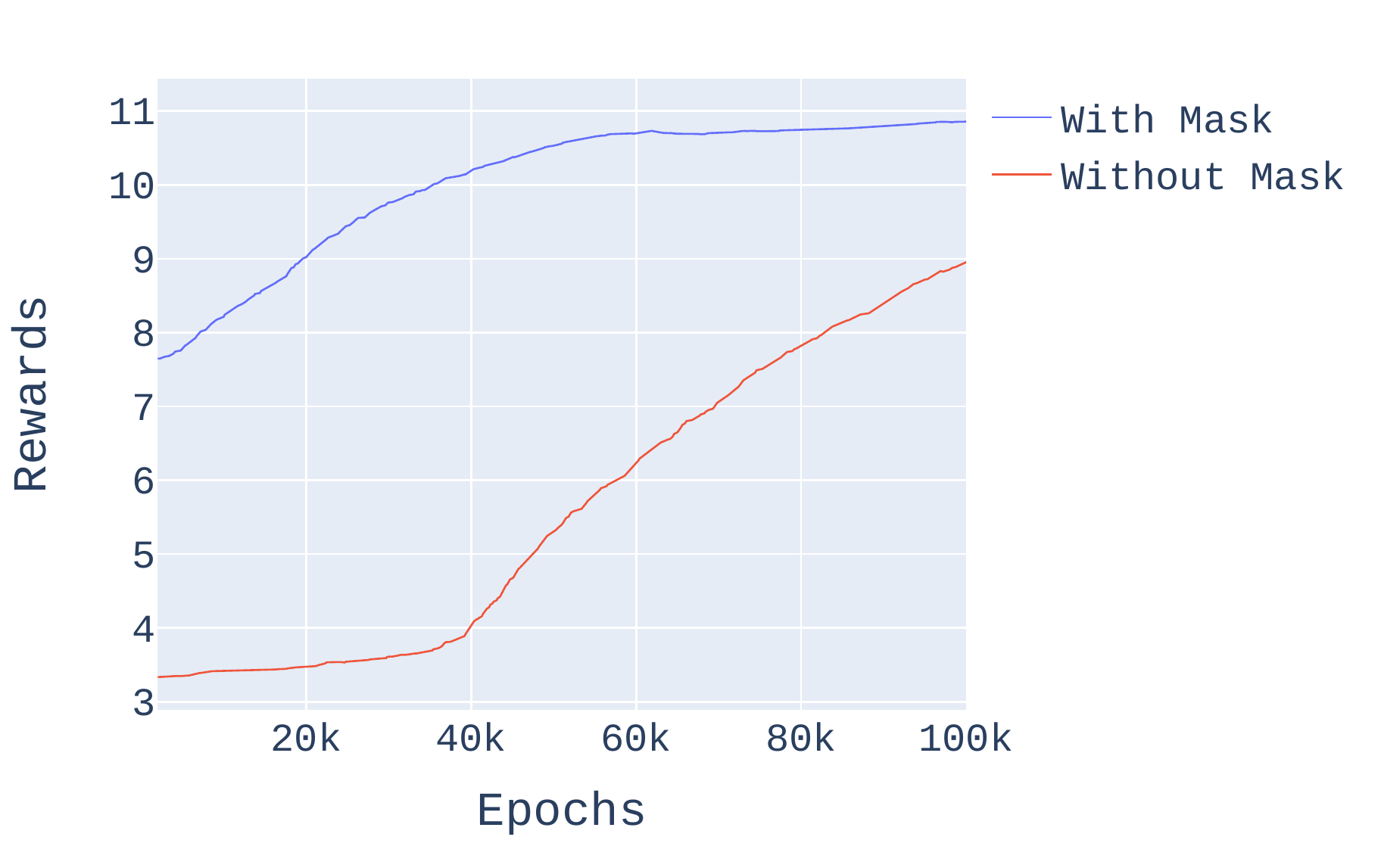}
  \caption{Reward comparison between node placement with output masking (blue) and without output masking (red).}
  \label{fig:mask_nomask}
\end{figure}

\begin{figure}[tb]
  \centering
  \includegraphics[width=\linewidth]{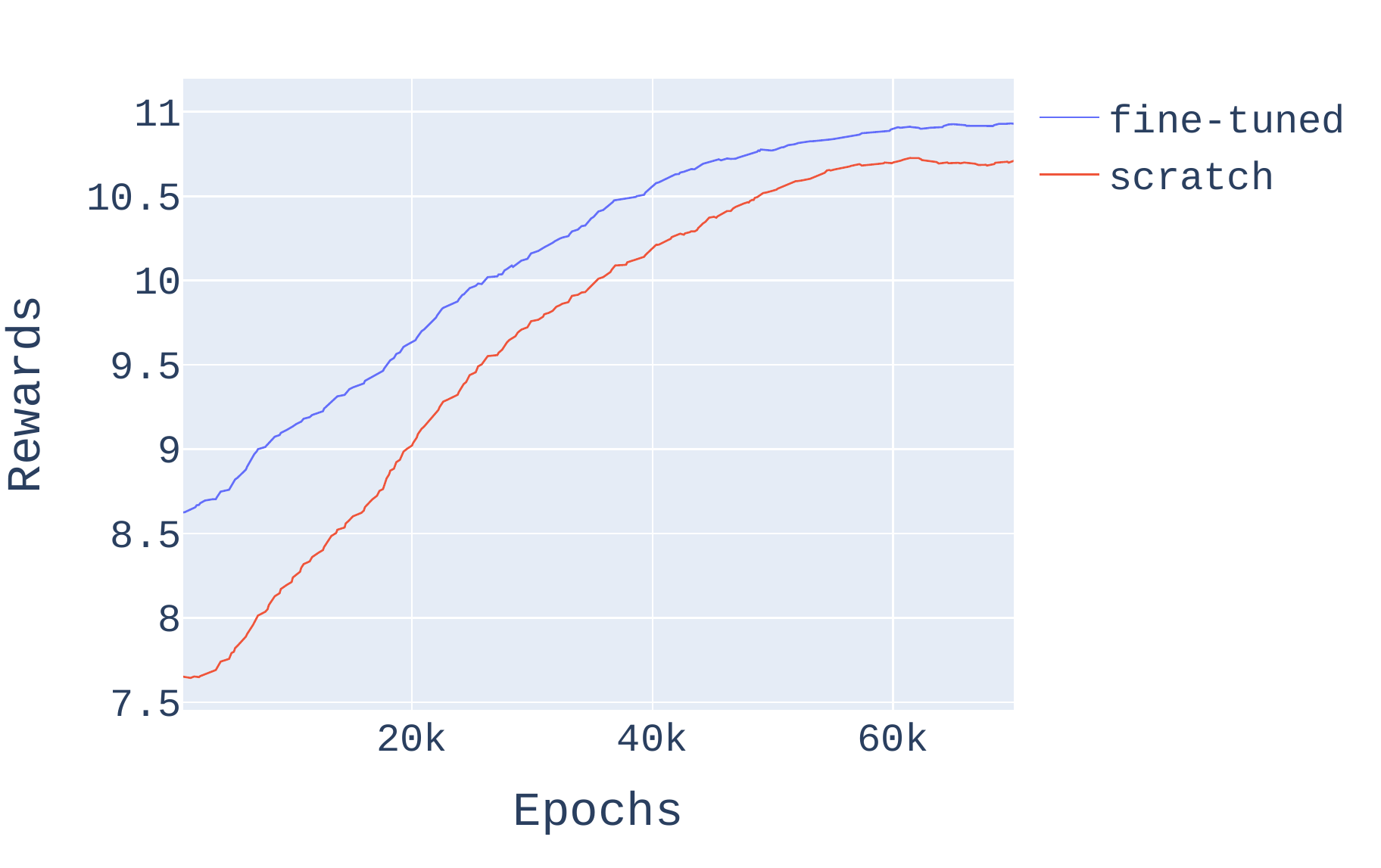}
  \caption{Comparison between training RL model from scratch (red) versus fine-tuning a model pre-trained on random graphs (blue). }
  \label{fig:pretrain_ifft}
\end{figure}

We also benchmarked our RL framework against a collection of random directed graphs meant to simulate real life applications.
We compare a baseline PPO method composed of three stacked MLPs against an actor model with the proposed GGA module.
We evaluate these models on varying application IR graphs with different complexities by increasing the total number of nodes.
We also tested on a larger device with 64 tiles.
In \figurename~\ref{fig:nodes_graph}, we observe that the RL approach finds mappings for a variety of IR graphs for randomly generated applications, and GGA improves the best schedule found for the same number of training epochs.
The GGA also finds an improved schedule in terms of clock cycles taken for a larger device configuration.
For the results in \figurename~\ref{fig:nodes_graph}, the epochs were limited to 50,000 for both models.
The cycle count is the number of SE execution steps taken to process all nodes for a mapping, which is calculated from the simulated SE environment.

\blue{In \figurename~\ref{fig:ifft_sa}, we evaluate the RL method for the FFT application and compare it against Simulated Annealing (SA) optimization method \cite{kirkpatrick1983optimization}.
We observe that our method finds higher rewards over time, while SA finds it difficult to improve reward in the SE's sparse search space.}

\subsection{GGA} \label{sec:GGA_result}

In \figurename~\ref{fig:ifft_rewards}, we see the addition of GGA module provides improved reward and sample efficiency for the FFT application by 10\%.
The node to be placed and SE state changes for each iteration over all nodes in the SDFs. The GGA module provides a representation of the entire IR graph while placing each node.

The attention module assists in highlighting the node dependencies when the RL model is predicting the placement of a node.
In \figurename~\ref{fig:ifft_attention}, we plot the attention matrix produced by the GGA module when mapping nodes from the FFT application shown in \figurename~\ref{fig:ifft_graph}.
The x-axis is the node index, the title indices which node is being placed and the color axis indicates attention scale.
We see that when placing node 1, the model focuses on nodes 2 and 3. When placing node 3, the attention is focused on node 6.
These attention values match our expectations since these nodes are related and have direct data dependencies.

\subsection{Node Iteration Order}

The iteration order is the sequence in which nodes are fed to the actor model in the RL framework and plays an important role in deciding whether the nodes can be successfully placed or not.
In \figurename~\ref{fig:ordered_placement}, we observe that iterating over nodes in topological ordering results in higher reward and eases the node placement task.
On the other hand, when placing the nodes randomly, sometimes nodes get placed before their predecessors are placed and the model needs to predict which tiles the predecessor nodes will be placed on, making training and the learning problem harder.

\subsection{Output Masking}
\label{subsec:output_masking}
The SE instruction scheduling task has several invalid actions for a given state that add noise to the samples and decision of the RL model, making convergence harder.
After placement of each node, the succeeding nodes have fewer placement options due to their data dependency with other nodes and device constraints.
Masking invalid placements reduces the search space as each node is being placed.
\figurename~\ref{fig:mask_nomask} demonstrates that masking improves the rewards obtained when masking is used by 20\% as compared to using a negative reward for invalid actions. We can also see that making helps improve sample efficiency of the RL model.

\subsection{Pre-training and fine-tuning}
After training the RL model on various applications, the RL model can be used for fine-tunning on a specific task or for inferencing.
In \figurename~\ref{fig:pretrain_ifft}, the RL model was pre-trained on a collection of randomly generated IR graphs and then fine-tuned for the FFT application.
The initial and final reward of the fine tuned RL model is higher than training the model from scratch.
This demonstrates that the model is able to reuse some of the previous experience of placing nodes for random input graphs.

\section{Conclusion}
\label{sec:conclusion}
The proposed RL mapper is a key component of the SE device toolchain.
It can search for optimal mappings while using the learning to map other previously unseen workloads more efficiently.
This improves upon the total time required to get mappings as compared to the existing manual placement approach and also reduces the amount of manual labor required.
It also allows for an automated search of mappings with different optimizations and trade-offs.
In this paper we have given a brief overview of the SE device architecture along with an analysis of how different design choices in terms of simulating the SE and neural network design impact the quality of mappings obtained. As future work we wish to explore:

\begin{itemize}
    \item Increasing sample efficiency of learning methods and improving the simulation environment for the SE by adding more constraints. 
    \item Integration of RL mapper into the SE toolset.
    \item Investigate the application of the proposed techniques for broader problems such as chip placement.
\end{itemize}

\section*{Acknowledgment}
\label{sec:acknowledgment}
The authors acknowledge Balint Fleischer, Glen Edwards, Jon Carter, Jeff Quigley, Mark Hur, Steve Pawlowski, and Tony Brewer for their continuous and extensive support for this project.

\bibliographystyle{IEEEtran}
\bibliography{all}

\end{document}